# Generation of Diffraction-Free Optical Beams Using Wrinkled Membranes

Ran Li[‡], Hui Yi[‡], Xiao Hu[‡], Leng Chen[‡], Guangsha Shi, Weimin Wang and Tian Yang[*]

*University of Michigan- Shanghai Jiao Tong University Joint Institute, National Key Lab of Nano/Micro Fabrication Technology, Key Lab for Thin Film and Microfabrication of the Ministry of Education, State Key Lab of Advanced Optical Communication Systems and Networks, Shanghai Jiao Tong University, Shanghai 200240, China*
[‡]*Equal contribution to this work*   [*]*tianyang@sjtu.edu.cn*

**Abstract:** We report the first demonstration of wrinkled membranes as a kind of optical focusing devices, which are low cost, light weight and flexible. Our device consists of concentric wrinkle rings on a gold-PDMS bilayer membrane, which converts collimated illuminations to diffraction-free focused beams. Beam diameters of 300-400 μm have been observed in the visible range. By comparing the theoretically calculated and experimentally measured focal spot profiles, we predict a focal spot size as small as around 50 μm if fabrication eccentricity can be eliminated.

## 1. Introduction

Wrinkling and buckling are commonplace phenomena in nature, e.g. skin wrinkles, finger prints, surface patterns of plants and folded mountains [1,2]. The most commonly employed model for wrinkling is a bilayer membrane which is composed of a rigid and thin elastic surface layer on top of a soft and thick elastic substrate. When the surface layer is under compressive strain by contraction of the substrate, the membrane spontaneously wrinkles to minimize its total energy which contains the bending energy and the stretching energy. Fabrication of periodic wrinkle patterns with desired patterns, periods and amplitudes have been theoretically modeled and experimentally developed in the last fifteen years or so [3-10]. The controllable and tunable formation of artificial wrinkle structures has attracted a lot of interest as a bottom-up technique for surface texture fabrication at the micrometer and nanometer scales. In the area of optics, wrinkle structures have been employed in applications including one dimensional gratings [4,11], microlens fabrication [12], flexible and tunable devices [13-16], light extraction for organic light emitting diodes [17], and light trapping for polymer photovoltaic devices [18].

In this paper, we report, as far as we know, the first demonstration of wrinkled membranes as optical focusing devices. Each device consists of concentric and uniformly periodic wrinkle rings in a gold(Au)-polydimethylsiloxane(PDMS) bilayer thin membrane. It has the same optical focusing and diffraction properties as an axicon lens, which converts collimated optical beams to diffraction-free focused beams in the focus region, which are Bessel beams in the ideal situation, and to rings in the farther space or the divergence region. We expect other focusing functions to be achieved in similar devices, which will be explained in Section 5. We also report a study on the limit of these bottom-up devices' focusing ability. In this study, by comparing the theoretically calculated focal spot profile and the experimentally measured one, we find that the devices reported in this paper are limited by a systematic eccentricity in fabrication, and predict that a focal spot size of as small as around 50 μm in the visible range will be achieved if eccentricity is eliminated.

## 2. Devices and the fabrication method

The whole fabrication procedure and a schematic illustration of the final device are shown in Fig. 1. First, we placed a piece of PDMS film, which was 170 μm thick, flatly on top of a Compact Disc (CD). Uncured PDMS mixture, in the form of viscous fluid, was applied between the CD and the PDMS film then oven cured to fix them together. Next, a radially oriented and uniformly distributed strain is applied on the film by pushing the center of the freely hanging PDMS film with a pointed bullet, stretching the film by 40%. To mitigate friction between the bullet and the PDMS layer, methanol was applied between them as a lubrication fluid. Then we coated a thin layer of Au onto the PDMS film, in a sputter coater which is commonly used for Scanning Electron Microscopy (SEM). Finally, we withdrew the bullet, and the Au-PDMS bilayer membrane wrinkled to reach the lowest mechanical energy.

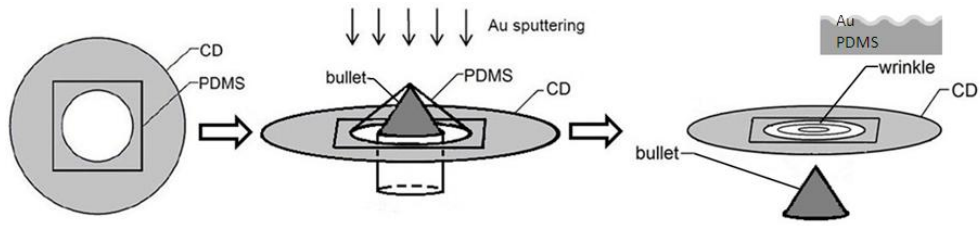

Fig. 1. Fabrication procedure for a Au-PDMS bilayer membrane with concentric wrinkle rings.

Fig. 2(a) shows the photo of a fabricated device under room light. The shiny colors on the membrane came from the diffraction of room light by the periodic wrinkles, while the central part of the membrane that had been in contact with the bullet looks opaque. As shown by the optical micrograph in Fig. 2(b), this device contains concentric wrinkle rings centered around the bullet pushing point, with a locally uniform period. The wrinkles form a regular and long-range ordered pattern over the 14 mm diameter CD hole, except for the part which had been in contact with the bullet as shown in Fig. 2(c).

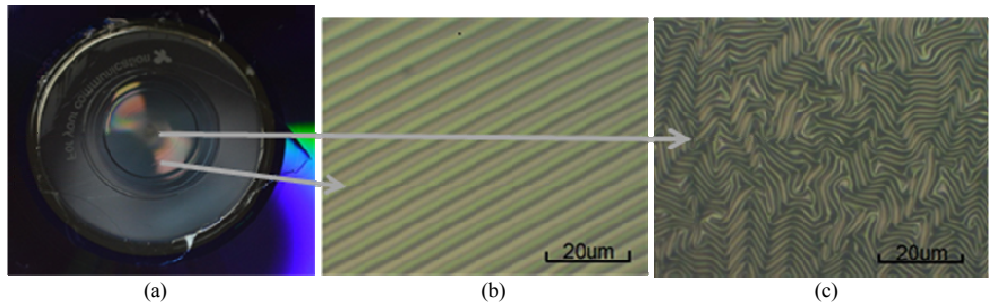

Fig. 2. A piece of Au-PDMS bilayer membrane with concentric wrinkle rings. (a) The whole membrane on a CD, observed under room light. (b) The concentric wrinkle rings under an optical microscope, which are oriented along the azimuthal direction. (c) The opaque central part of the membrane under an optical microscope.

## 3. Optical focusing performance

We characterized the focusing performance of the device in Fig. 2 by illuminating its wrinkle side with a collimated beam of light at normal incidence. The illuminating beam was from a broadband halogen lamp and 11 mm in diameter. A schematic illustration of the experiment is plotted in Fig. 3(a). It shows that the +1st order diffraction off the concentric and uniformly periodic wrinkle rings is analogous to the refraction from the conical surface of an axicon, whose interference with itself produces a diffraction-free focused beam in the focus region of the device. In the divergence region, the +1st and -1st orders of diffraction form two optical rings respectively, separated by the opaque central part of the membrane. In Fig. 3(b), we took a side view of the focused optical beam with a piece of white paper, which is a five centimeter long, 300-400 μm in diameter, and diffraction-free rainbow beam.

The optical characterization results have been compared with theoretical modeling. The wrinkled membrane in this device has an average period around 4.7 μm and a Au layer thickness around 13 nm. In theory, a concentric ring grating (with a good central part) that has a 4.7 μm period will focus a 11 mm diameter and 0.5 μm wavelength (blue) incident beam to a 5.2 cm long line, and a 0.7 μm wavelength (red) incident beam to a 3.7 cm long line. This is in good agreement with the experimentally observed rainbow line shown in Fig. 3(b). The amplitude of wrinkles determines the wavelength where diffraction is most efficient. The peak-to-peak amplitude of the wrinkles in this device is around 0.7 μm, according to Atomic Force Microscopy (AFM) measurement. A finite difference time domain (FDTD) calculation for a one dimensional sinusoidal grating with the same material, period, and amplitude as the wrinkle-ring device shows a maximum +1st order diffraction efficiency for blue light of any polarizations, which is around 24%. By replacing gold with transparent surface layer materials, e.g. oxidized PDMS [4], the diffraction efficiency will be further increased to 34% according to FDTD calculation. The focal depth and peak efficiency wavelength can be tuned straightforwardly by tuning the period and amplitude of the wrinkles. The period scales linearly with the thickness of Au film, while the amplitude-to-period ratio scales with the square root of compressive strain [5].

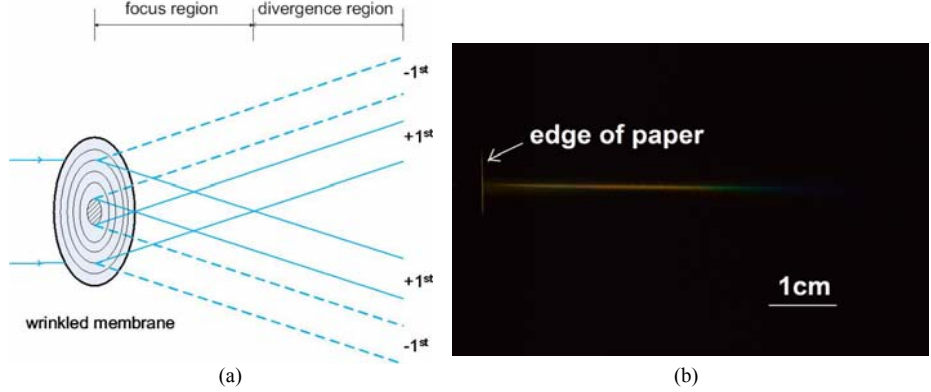

Fig. 3. Optical characterization of the focusing performance of the wrinkled membrane. (a) A schematic of the optical experiment. (b) A side view of the diffraction-free beam in the focus region, on a piece of white paper. The front edge of the paper is 1 mm from the membrane.

**4. Period eccentricity and the limit of focusing ability**

Just as many other bottom-up processes, wrinkle structures contain defects and non-uniformity, which are analogous to lens aberrations and result in focal spots larger than the diffraction limit. In this section, we report a study on this problem. We find that the microscopic local defects and non-uniformity only impose a limit on the focal spot size which is as small as around 50 μm, while the macroscopic period eccentricity is the dominant reason for the large beam diameters in our experiments.

It has been observed for all of our wrinkle-ring devices that the period is different at different azimuthal angles, which we call "period eccentricity" in this paper. Period eccentricity has been confirmed by both microscopy and by measuring the optical diffraction rings in the divergence region. The diffraction ring pattern renders a clear quantitative description of period eccentricity and is discussed as follows. In this experiment, we illuminated the same device as shown in Fig. 2 with a collimated beam from a He-Ne laser at normal incidence, which was around 12 mm in beam diameter and 633 nm in wavelength ($\lambda$). Fig. 4 shows a front view of the diffraction ring pattern at a distance of $z$=33 cm from the membrane. The inner ring comes from the +1st order diffraction, the outer ring comes from the -1st order diffraction, and the dark area between the rings comes from the opaque central part of the membrane. As illustrated in Fig. 3(a), each segment of wrinkle rings at a certain azimuthal angle, $\theta$, diffracts the incident laser beam to the $\theta$ point on the -1st order diffraction ring, and to the $\theta+\pi$ point on the +1st order diffraction ring. Therefore, the period of the wrinkle rings, $L$, at a certain azimuthal angle, $\theta$, can be calculated from the distance between the two corresponding diffraction points, $d$, by

$$L(\theta) = \frac{\lambda}{\sin\alpha}, \alpha = \arctan\left(\frac{d}{2z}\right). \quad (1)$$

The calculated $L\sim\theta$ diagram is plotted in Fig. 4(b). For example, by the lengths $d_1$ and $d_2$ in Fig. 4(a), we calculated the period between the pushing point and the top CD hole edge to be

4.5 um, where $\theta$=0, and the period between the pushing point and the bottom CD hole edge to be 5.0 um, where $\theta$=$\pi$. As an aside, the different diffraction properties at different azimuthal angles could be compared to coma, which is the worst aberration effect of glass lenses.

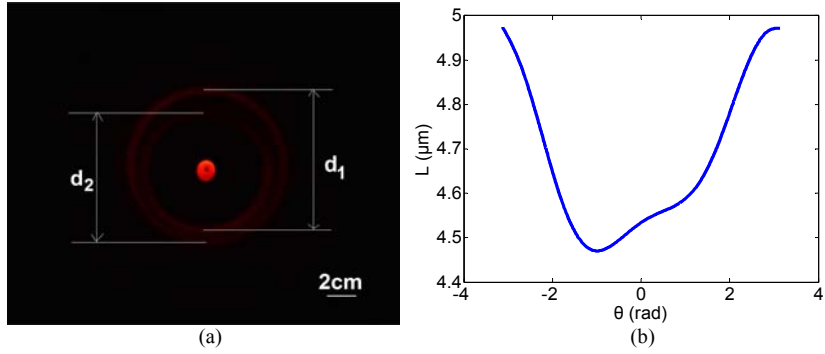

(a) (b)

Fig. 4. (a) Illuminating a wrinkled membrane with a He-Ne laser beam at normal incidence, a pair of rings at a distance of 33cm is observed on a piece of white paper. The bright spot in the center is non-diffracted laser light. (b) Calculated wrinkle period at each azimuthal angle.

Then we calculated the focal spot profile assuming that period eccentricity $L(\theta)$, as depicted in Fig. 4(b), is the only imperfectness of the device. The same He-Ne laser incidence scheme as above was used. The wrinkled membrane's electric field amplitude transmittance function, $U_t$, was taken to be the following:

$$U_t = circ(\frac{\rho}{R})e^{j\frac{(n-1)kA}{2}\sin(\frac{2\pi}{L(\theta)}\rho)}, \qquad (2)$$

where $\rho$ is the radial coordinate, $R$ is the radius of the membrane, $n$=1.4 is the refractive index of PDMS, $A$=0.7 μm is the peak-to-peak amplitude of the wrinkles, and $k = \frac{2\pi}{\lambda}$ is the wave vector. Here we have assumed that the wrinkles have a sinusoidal shape, that $U_t$ has a uniform amplitude of 1, that $U_t$'s phase changes from $-\frac{(n-1)kA}{2}$ to $\frac{(n-1)kA}{2}$ in each wrinkle period, and that $U_t$'s phase is 0 at the center point regardless of the value of $\theta$. Then the scalar electric field on the observation point $(x,y,z)$ was calculated by the Huygens–Fresnel principle:

$$U(x,y,z) = \frac{z}{j\lambda} \iint U_t(\xi,\eta) \frac{e^{jkr}}{r^2} d\xi d\eta, \qquad (3)$$

where ($\xi,\eta$) are Cartesian coordinates on the back plane of the device ($z$=0), and $r = \sqrt{z^2 + (x-\xi)^2 + (y-\eta)^2}$. Finally, the electric field intensity profile on the observation plane was plotted as $|U(x,y,z_0)|^2$, as shown in Fig. 5(a), where $z_0$=3 cm. One point per 2 μm×2 μm area on the observation plane has been calculated, and the nearest 3×3 points have been averaged. The calculated focal spot at a distance of 3 cm from the device is 300-400 μm in size, with rhombus-shaped sides and other line-shaped features.

An experimental image of the focal spot taken by direct illumination onto a monochrome CMOS camera chip is shown in Fig. 5(b), with the calculated features overlapping on top for comparison. Each pixel of the CMOS chip is a 5.2 μm×5.2 μm square. The calculation and the experimental image match each other well, except that the line-shaped features, which result from interference, are less sharper in the experimental image, which are around 50 μm. This nice comparison indicates that period eccentricity dominantly widens and shapes the focused beam to what have been observed in experiments. The 50 μm feature lines represent a focusing limit which is a combined consequence of the microscopic local defects, non-uniformity and other macroscopic errors except period eccentricity.

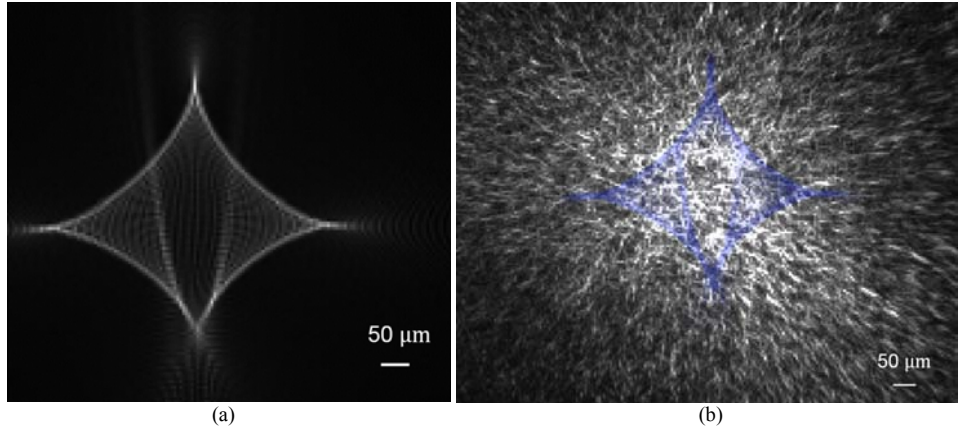

(a)                            (b)

Fig. 5. Calculated and measured He-Ne laser focal spot profiles off a wrinkled membrane. (a) Calculated electric field intensity profile at a distance of 3 cm from the membrane. (b) An image of the focal spot taken by direct illumination onto a monochrome CMOS camera chip, with the calculated features from (a) overlapping on top in blue. The distance between the CMOS chip and the membrane is 3-3.5 cm.

## 5. Discussion and conclusion

The significant period eccentricity in our devices is most probably due to a non-uniform Au layer thickness, the reason of which is still under exploration. If we can eliminate this effect by improving our fabrication, we expect to obtain focal spots as small as around 50 μm, which will find many applications such as optical trapping, optical coherence tomography, telescopes, laser manufacturing, etc. [19-21] In addition, we expect other types of wrinkled membrane focusing devices, e.g. Fresnel lenses, to be fabricated by further strain distribution engineering. This can be done by placing a circular shutter centered above the membrane during Au deposition, so that the thickness of Au film changes as designed, and consequently desired $L(\rho)$ profiles can be achieved. Such wrinkled membrane focusing devices have advantages over bulky glass optics and lithographically patterned zone plates in terms of low cost, light weight, flexibility and suitability for large area fabrication.

In conclusion, we have reported, as far as we know, the first demonstration of optical focusing using wrinkled membranes. A concentric wrinkle ring pattern was formed in a 13 nm Au – 170 μm PDMS bilayer membrane, with a period around 4.7 μm. The device focuses visible collimated illuminations, which have diameters of around 1 cm, to diffraction-free beams which have diameters of 300-400 μm and focal lengths of 4-5 cm. Further, by studying the diffraction rings and comparing the calculated and experimentally measured focal spot profiles, we have identified period eccentricity as the dominant beam widening and shaping factor for our current devices. If device eccentricity can be eliminated, we predict the focal spot to be as small as around 50 μm. Therefore we anticipate the wrinkled membranes to become a viable solution where low cost, light weight and flexible focusing lenses are in need.

## Acknowledgements

This work is supported by the Shanghai Pujiang Program under grant #10PJ1405300, the National Science Foundation of China under grant #61275168, and the National High Technology Research and Development Program of China (863 Program) under grant #2011AA050518. Li, Chen, Shi and Wang fabricated the devices. Yi and Yang performed the optical measurements. Hu finished the calculations. Yang led the reported work and prepared the manuscript. We thank Prof. Qihui Shi and Dr. Yuliang Deng from the Shanghai Center for Systems Biomedicine for assistance with fabrication.